\documentclass[aps,prl,showpacs,twocolumn,amsmath,amssymb]{revtex4-1}
\usepackage{graphicx,bbm}

\newcommand{\be}{\begin{equation}}
\newcommand{\ee}{\end{equation}}
\newcommand{\un}[1]{{\underline{#1}}}
\newcommand{\bra}[1]{{\langle #1 \vert}}

\newcommand{\ket}[1]{{\vert #1 \rangle}}

\newcommand{\ave}[1]{{\langle #1\rangle}}
\newcommand{\ii}{ {\rm i} }
\newcommand{\dd}{ {\rm d} }

\newcommand{\RaR}{\mathbb{R}}
\newcommand{\CC}{\mathbb{C}}

\newcommand{\mm}[1]{{\mathbf{#1}}}

\def\tr{{{\rm tr}}}

\def\one{\mathbbm{1}}

\begin{document}

\title{Exact nonequilibrium steady state of a strongly driven open XXZ chain}
\author{Toma\v{z} Prosen}
\affiliation{Department of Physics, FMF,  University of Ljubljana, Jadranska 19, 1000 Ljubljana, Slovenia}


\date{\today}

\begin{abstract}
An exact and explicit ladder-tensor-network ansatz is presented for nonequilibrium steady state of an anisotropic Heisenberg XXZ spin-1/2 chain which is driven far from equilibrium with a pair of Lindblad operators acting on the edges of the chain only. We show that the steady-state density operator of a finite system of size $n$ is -- apart from a normalization constant -- a polynomial of degree $2n-2$ in the coupling constant.
Efficient computation of physical observables is facilitated in terms of a transfer-operator reminiscent of a classical Markov process. In the isotropic case we find cosine spin profiles, $1/n^2$ scaling of the spin current, and long-range correlations in the steady state.
This is a fully nonperturbative extension of a recent result [Phys.~Rev.~Lett.~{\bf 106},~217206~(2011)].
\end{abstract}

 
\maketitle

{\em Introduction.-}
The Heisenberg model \cite{heisenberg} of coupled quantum spins 1/2 is the oldest many body quantum model of strong interactions. In spite of being extremely simple it exhibits (in particular its anisotropic version, the XXZ model) a rich variety of equilibrium and nonequilibrium physical behaviors. In nature it provides an excellent description of the so called spin-chain materials \cite{spinchain}, and it is believed to provide the key for understanding of various collective quantum phenomena in low dimensional strongly interacting systems, such as magnetic or superconducting transitions in two dimensions. Although equilibrium (thermodynamic) properties of XXZ chain are well understood in terms of Bethe Ansatz (BA) \cite{takahashi}, as the model represents a paradigmatic example of quantum integrable systems, its nonequilibrium properties at finite temperature are subject to lively debate \cite{affleck}. 

Ground states of strongly correlated systems generically satisfy area laws \cite{jens} for block entropy characterizing bipartite quantum entanglement, so they can be efficiently described by the so called matrix product states (MPS) or more general tensor networks \cite{frank}. MPS of small rank can provide even exact description of ground states, say in valence bond solids exemplified by the famous AKLT model \cite{aklt}. 
In fact, even BA eigenfunctions can be written in terms of MPS \cite{hosho}.
On the other hand, using the approach of open quantum systems and Markovian master equations \cite{breuer}, nonequilibrium steady states (NESS) of large one-dimensional locally interacting and dissipationless quantum systems put between a pair of unequal macroscopic reservoirs \cite{saito,wichterich}, can be described in terms of a fixed point, or `ground state' in the Liouville space, for a Hermitian super-operator with non-Hermitian boundary terms \cite{3Q}. Application of the density matrix renormalization group (DMRG) for simulation of such problems showed that a sort of super-area law is generically valid, and the density operator of NESS can be well described by a matrix product operator (MPO) of low rank \cite{pz}. However, no far-from-eqilibrium analogues of AKLT model have been known so far, and the purpose of this Letter is to show an explicit construction of an exact MPO form of NESS for a boundary driven XXZ spin chain. More precisely, a matrix element of the many-body density operator is a contraction of a very appealing ladder tensor network (LTN). 
  
We have recently proposed a new method \cite{new} to solve for a Liouvillian fixed point of the XXZ chain, perturbatively in the system-bath coupling constant. This method which expresses NESS in the form of a MPO with near-diagonal infinite rank matrices -- reminiscent of a classical Markov process in the auxiliary space --  suggests new ways of integrability of strongly 
nonequilibrium quantum lattice gasses and appears to be unrelated to BA. In this Letter we show that -- quite nontrivially -- a fully nonperturbative extension of this method exists (in the strong driving limit of maximal bias, $\mu=1$ in notation of Ref.~\cite{new}), with the constituent matrices satisfying the same cubic matrix algebra (essentially different from quadratic algebras characterizing exactly solvable classical probabilistic lattice gasses, the so-called exclusion processes \cite{blythe}), but with modified boundary relations. From our exact analysis, we: (i) prove ballistic transport (size $n$ independent spin current) in the easy-plane regime,  (ii) derive coupling independent cosine spin-profiles, $1/n^2$ scaling of the spin current and long-range spin-spin correlations in the isotropic regime, and (iii) prove insulating behavior in the easy-axis regime with kink-shaped spin profiles and exponentially  (in $n$) decaying currents. We note that the physics of {\em near-equilibrium} XXZ chain is essentially different. There one has perturbative and numerical evidence of spin diffusion \cite{markonew,robin} in the easy-axis regime, and alternative super-diffusive anomalous scaling in the isotropic point \cite{markonew} indicating very rich phenomenology of the model.

{\em Nonequilibrium steady state.-} We consider the Markovian master equation in the Lindblad form \cite{breuer,wichterich}
\be
\frac{\dd \rho(t)}{\dd t} = -\ii [H,\rho(t)] + \sum_k 2 L_k \rho(t) L^\dagger_k - \{ L^\dagger_k L_k,\rho(t)\}
\label{eq:lindblad}
\ee
for an open XXZ spin 1/2 chain with the Hamiltonian
$
H = \sum_{j=1}^{n-1}h_j,$
$h_j:=2\sigma^+_j \sigma^-_{j+1} + 2\sigma^-_j \sigma^+_{j+1} + \Delta \sigma^{\rm z}_j \sigma^{\rm z}_{j+1}$
and symmetric Lindblad driving of {\em coupling strength} $\varepsilon$  acting on the edges of the chain only
$
L_1 = \sqrt{\varepsilon}\sigma^+_1, \quad L_2 = \sqrt{\varepsilon}\sigma^-_n.
$
We write Pauli operators on a tensor product space ${\cal F}_n=(\CC^2)^{\otimes n}$, as
$\sigma^s_j = \one_{2^{j-1}}\otimes \sigma^s \otimes \one_{2^{n-j}}$, $\one_d$ being a $d$-dimensional unit matrix, where
$\sigma^\pm = \frac{1}{2}(\sigma^{\rm x} \pm\ii \sigma^{\rm y})$ and $\sigma^{\rm x,y,z}$ are the standard Pauli matrices.
 
NESS is a fixed point of the flow (\ref{eq:lindblad}) $\rho_\infty = \lim_{t\to\infty}\rho(t)$ 
\be
-\ii [H, \rho_\infty] + \varepsilon \hat{\cal D} \rho_\infty = 0,
\label{eq:fixedpoint}
\ee 
with the dissipator map
\be 
\hat{\cal D} \rho :=
2 \sigma^+_1 \rho \sigma^-_1 - \{\sigma^-_1 \sigma^+_1,\rho\}+
2 \sigma^-_n \rho \sigma^+_n - \{\sigma^+_n \sigma^-_n,\rho\}. 
\label{eq:dissipator}
\ee
The quantum magnetic transport model (\ref{eq:lindblad}) can be derived \cite{giuliano} by using standard assumptions \cite{breuer}, or alternatively, from an 
exact microscopic protocol of {\em repeated interactions} \cite{karevski,hartmann} where the left/right boundary spins are repeatedly and frequently put into arbitrarily strong contact with fresh up/down polarized magnets \cite{temperature}.
We shall now construct an explicit form of $\rho_\infty$ in terms of the LTN ansatz, or equivalently, in terms of a product of two MPOs,  which is exact for any value of the coupling parameter $\varepsilon$. In fact, our simple explicit form allows us to study analytic dependence of NESS on $\varepsilon$.
 
 \noindent
{\bf Theorem.} The normalized fixed-point solution of Eq. (\ref{eq:fixedpoint}) reads $\rho_\infty = (\tr R)^{-1} R$ with
\be
R= S_n S^\dagger_n
\label{eq:defR}
\ee
and $S_n$ a non-Hermitian {\em matrix product operator}
\be 
S_n = \!\!\!\!\!\!\!\!\!\!\!\!\!\sum_{(s_1,\ldots,s_n)\in\{+,-,0\}^n}\!\!\!\!\!\!\!\!\!\!\!\!\!\!\bra{0} \mm{A}_{s_1}\mm{A}_{s_2}\cdots\mm{A}_{s_n}\ket{0} \sigma^{s_1}\otimes \sigma^{s_2} \cdots\otimes \sigma^{s_n}
\label{eq:MPS}
\ee
where $\sigma^{0} \equiv \one_2$ and $\mm{A}_0, \mm{A}_\pm$ is a triple of near-diagonal matrix operators acting on an infinite-dimensional auxiliary Hilbert space
${\cal H}$ spanned by an ortho-normal basis $\{ \ket{0},\ket{1},\ket{2},\ldots \}$:
\begin{eqnarray}
\mm{A}_0 &=& \ket{0}\bra{0} + \sum_{r=1}^\infty a^0_r \ket{r}\bra{r}, \nonumber \\
\mm{A}_+ &=& \ii \varepsilon \ket{0}\bra{1} + \sum_{r=1}^\infty a^+_r \ket{r}\bra{r\!+\!1},\label{eq:explicitA}\\
 \mm{A}_- &=& \ket{1}\bra{0} + \sum_{r=1}^\infty a^-_r \ket{r\!+\!1}\bra{r}, \nonumber
 \end{eqnarray}
with matrix elements (writing $\lambda := \arccos \Delta \in \RaR \cup \ii\RaR$)
\begin{eqnarray}
a^0_r        &=& \cos\left(r\lambda\right) + \ii\varepsilon \frac{ \sin\left(r \lambda \right)}{2\sin \lambda}, \nonumber\\
a^+_{2k-1} &=& c\sin\left(2k\lambda\right)+ \ii\varepsilon \frac{c \sin\left((2k\!-\!1)\lambda\right)\sin\left(2k\lambda\right)}{2(\cos\left((2k\!-\!1)\lambda\right)+\tau_{2k-1})\sin\lambda},  \nonumber \\
a^+_{2k}     &=&  c\sin\left(2k\lambda\right)- \ii\varepsilon \frac{c (\cos\left(2k\lambda\right)+\tau_{2k})}{2\sin\lambda}, \label{eq:a} \\
a^-_{2k-1}  &=& -\frac{\sin\left((2k\!-\!1)\lambda\right)}{c} + \ii\varepsilon \frac{\cos\left((2k\!-\!1)\lambda\right)+\tau_{2k-1}}{2c\sin\lambda}, \nonumber \\
a^-_{2k}      &=& -\frac{\sin\left((2k\!+\!1)\lambda\right)}{c} - \ii\varepsilon \frac{ \sin\left(2k\lambda\right)\sin\left((2k\!+\!1)\lambda\right)}{2c(\cos\left(2k\lambda\right)+\tau_{2k})\sin\lambda}. \nonumber 
\end{eqnarray}
Constant $c \in \CC - \{0\}$ and signs $\tau_r \in \{\pm 1\}$ are arbitrary,
i.e. all choices of $c,\tau_r$ give identical operator $S_n$ (\ref{eq:MPS}).

\medskip
\noindent
{\bf Proof.} We start by showing the following useful identity
\be
[H,S_n] = -\ii \varepsilon (\sigma^{\rm z}\otimes S_{n-1} - S_{n-1}\otimes \sigma^{\rm z}).
\label{eq:HS}
\ee
It is important to observe that the ansatz (\ref{eq:MPS}) does not contain any $\sigma^{\rm z}_j$ operator, while $[H,S_n]$ can contain only terms with a single $\sigma^{\rm z}_j$.  
Eq.  (\ref{eq:HS}) implies that all the terms of $[H,S_n]$ where $\sigma^{\rm z}_j$ appear in the bulk
 $1 < j < n$ should vanish, resulting in exactly the same argument as in \cite{new} leading to the same eight 3-point algebraic conditions:
 \begin{eqnarray}
\!&& [\mm{A}_0,\mm{A}_\pm\mm{A}_\mp] = 0, \;\;\,\qquad\qquad \{ \mm{A}_0,\mm{A}^2_\pm\} = 2\Delta \mm{A}_\pm\mm{A}_0\mm{A}_\pm , \nonumber \\
\!&& 2\Delta\{\mm{A}^2_0,\mm{A}_\pm\} - 4\mm{A}_0\mm{A}_\pm\mm{A}_0 = \{\mm{A}_\mp,\mm{A}^2_\pm\} - 2\mm{A}_\pm\mm{A}_\mp\mm{A}_\pm , \nonumber \\
\!&& 2\Delta [\mm{A}_0^2,\mm{A}_\pm]  = [\mm{A}_\mp,\mm{A}_\pm^2] .  \label{eq:algebra}
\end{eqnarray}
However, the boundary conditions should be different as in the perturbative case \cite{new}.
Namely the remaining set of terms where $\sigma^{\rm z}_j$ appears at $j=1$ or $j=n$ in $[H,S_n]$ is reproduced exactly by the right-hand-side of (\ref{eq:HS}) if
the following additional algebraic conditions are satisfied
\begin{eqnarray}
\bra{0}\mm{A}_-&=& \bra{0}\mm{A}_+(\mm{A}_-\mm{A}_+ -\ii\varepsilon\one) = \bra{0} \mm{A}_+\mm{A}_-^2 = 0,\nonumber \\
\mm{A}_+\ket{0} &=& (\mm{A}_-\mm{A}_+- \ii\varepsilon\one)\mm{A}_-\ket{0} = \mm{A}_+^2\mm{A}_-\ket{0} = 0, \nonumber \\
\bra{0}\mm{A}_0 &=& \bra{0}, \quad \mm{A}_0\ket{0} = \ket{0}, \quad \bra{0}\mm{A}_+\mm{A}_-\ket{0} = \ii\varepsilon.\qquad \label{eq:boundary}
\end{eqnarray}
Note the simple extra terms with amplitude $-\ii \varepsilon$ in comparison to Eqs. (12) of Ref.~\cite{new}. Indeed, in order to get, e.g. a term with $\sigma^{\rm_z}_1$ in $[h_1,S_n]$, $s_1$ in (\ref{eq:MPS}) has to be $+$
and then the condition $\bra{0}\mm{A}_+(\mm{A}_-\mm{A}_+ -\ii\varepsilon\one)=0$ ensures that exactly $S_{n-1}$ will be constructed on the sites $(2,\ldots,n)$.

Verifying (\ref{eq:algebra}) and (\ref{eq:boundary}), which imply (\ref{eq:HS}), for the representation (\ref{eq:explicitA},\ref{eq:a}) results in trigonometric identities.

The rest of the proof is to show that (\ref{eq:HS}) implies (\ref{eq:fixedpoint}), or $\ii [H,R] =  \varepsilon \hat{\cal D}R$ (a). Left-hand-side of (a) can be transformed to
$[\ii H,S_n]S_n^\dagger + S_n[\ii H,S_n]^\dagger = \varepsilon \{
S_n (\sigma^{\rm z}\otimes S_{n-1}^\dagger) - S_n (S_{n-1}^\dagger\otimes \sigma^{\rm z})  +  (\sigma^{\rm z}\otimes  S_{n-1}) S_n^\dagger - (S_{n-1}\otimes \sigma^{\rm z}) S_n^\dagger\}$ (b).
Equation (\ref{eq:boundary}) implies that the first left-most (right-most) nontrivial operator of every term of $S_n$ is $\sigma^+$ ($\sigma^-$).
Thus we write $S_n =: \sigma^0 \otimes S_{n-1} + \sigma^+ \otimes P_{n-1} =: S_{n-1}\otimes \sigma^0 + Q_{n-1} \otimes \sigma^-$, 
defining $Q_{n-1}$ and $P_{n-1}$ as operators over ${\cal F}_{n-1}$, so the expression (b) further equals
$\varepsilon\{
2\sigma^{\rm z}\otimes S_{n-1}S_{n-1}^\dagger - \sigma^+ \otimes P_{n-1}S_{n-1}^\dagger - \sigma^- \otimes S_{n-1} P_{n-1}^\dagger -
2S_{n-1}S_{n-1}^\dagger \otimes \sigma^{\rm z} - Q_{n-1}S_{n-1}^\dagger \otimes \sigma^- - S_{n-1}Q_{n-1}^\dagger \otimes \sigma^+\}
$ (c). On the other hand, writing the dissipator (\ref{eq:dissipator}) as a sum of two local terms $\hat{\cal D}=\hat{\cal D}_{\rm L}\otimes \hat{\one}_{n-1}+
\hat{\one}_{n-1}\otimes \hat{\cal D}_{\rm R}$,
we have for the right-hand-side of (a): 
$(\varepsilon \hat{\cal D}_{\rm L}\otimes \hat{\one}_{n-1})(S_n S_n^\dagger)+
(\hat{\one}_{n-1}\otimes \varepsilon\hat{\cal D}_{\rm R})(S_n S_n^\dagger)$ (d). 
The first term of (d) can further be written out
as $(\varepsilon\hat{\cal D}_{\rm L}\otimes \hat{\one}_{n-1})
[(\sigma^0 \otimes S_{n-1} + \sigma^+\otimes P_{n-1})
(\sigma^0 \otimes S^\dagger_{n-1} + \sigma^-\otimes P^\dagger_{n-1})]=
\varepsilon\hat{\cal D}_{\rm L}(\sigma^0) \otimes S_{n-1} S^\dagger_{n-1} + \varepsilon\hat{\cal D}_{\rm L}(\sigma^-)\otimes S_{n-1} P^\dagger_{n-1} + 
\varepsilon\hat{\cal D}_{\rm L}(\sigma^+)\otimes P_{n-1} S^\dagger_{n-1} + \varepsilon\hat{\cal D}_{\rm L}(\sigma^+\sigma^-)\otimes P_{n-1}P^\dagger_{n-1}$.
Since, 
$\hat{\cal D}_{\rm L}(\sigma^0) = 2\sigma^{\rm z}$, 
$\hat{\cal D}_{\rm L}(\sigma^\pm) = -\sigma^\pm$, 
$\hat{\cal D}_{\rm L}(\sigma^+\sigma^-) = 0$, we arrive at exactly the first three terms of (c).
In an analogous way the second term of (d) results in the last three terms of (c). QED

\begin{figure}
         \centering	
	\includegraphics[width=0.6\columnwidth]{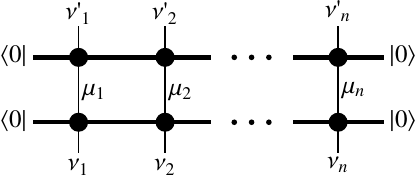}
	\caption{LTN contracting to a NESS density matrix element (\ref{eq:LTN}). Thin (thick) lines represent bond dimension $2$ ($d$).} 	\label{fig:1}
\end{figure}

{\em Corollaries.-} Let us now derive some implications of our ansatz (\ref{eq:defR},\ref{eq:MPS}): (i) Let $\ket{\un{\nu}}$, $\un{\nu}=(\nu_1,\nu_2,\ldots,\nu_n)\in\{0,1\}^n$ denote the canonical many-body basis of 
${\cal F}_n$, $\sigma^{\rm z}_j \ket{\un{\nu}}=(1-2\nu_j)\ket{\un{\nu}}$. Then the matrix elements of MPO (\ref{eq:MPS}) can be written out as ($\pm\equiv\pm 1$)
\be
\bra{\un{\nu}'}S_n\ket{\un{\nu}} = \bra{0}\mm{A}_{\nu_1-\nu'_1}\mm{A}_{\nu_2-\nu'_2} \cdots \mm{A}_{\nu_n-\nu'_n}\ket{0}.
\label{eq:matS}
\ee
(ii) This many-body matrix is {\em upper triangular}, i.e. $\bra{\un{\nu}'}S_n\ket{\un{\nu}} =0$ if ${\rm seq}(\un{\nu}') > {\rm seq}(\un{\nu})$  
(where ${\rm seq}(\un{\nu}):=\sum_{j=1}^n \nu_j 2^{n-j}$) following from $\bra{0}\mm{A}_-=0$ (\ref{eq:boundary}) hence Eq.~(\ref{eq:defR}) is 
{\em the Cholesky decomposition} of the many-body density matrix. We also have {\em unit diagonal} $\bra{\un{\nu}}S_n\ket{\un{\nu}}=1$, following from $\bra{0}\mm{A}_0=\bra{0}$ (\ref{eq:boundary}),
implying that NESS is  always of {\em full rank}.
(iii) Inserting the identity $\one=\sum_{\un{\mu}} \ket{\un{\mu}}\bra{\un{\mu}}$ into (\ref{eq:defR}), the matrix elements of density operator are obtained via contraction of a LTN 
(Fig.~\ref{fig:1})

\begin{eqnarray}
\!\!\!\!\!\!\!\bra{\un{\nu}'}R\ket{\un{\nu}} = \!\!\!\sum_{\un{\mu}\in\{0,1\}^n} 
&&\bra{0}\mm{A}_{\mu_1-\nu'_1}\mm{A}_{\mu_2-\nu'_2} \cdots \mm{A}_{\mu_n-\nu'_n}\ket{0} \nonumber \\
\times &&\bra{0}\bar{\mm{A}}_{\mu_1-\nu_1}\bar{\mm{A}}_{\mu_2-\nu_2} \cdots \bar{\mm{A}}_{\mu_n-\nu_n}\ket{0}.
\label{eq:LTN}
\end{eqnarray}
$\bar{\mm{A}}_s$ denote the complex-conjugate matrices, obtained from (\ref{eq:explicitA}) by complex-conjugating the amplitudes (\ref{eq:a}), equivalent to flipping the sign of $\varepsilon$, $\bar{\mm{A}}_s = \mm{A}_s|_{-\varepsilon}$ for a suitably chosen $c$ (say as in Ref. \cite{new}).
(iv) As the matrices (\ref{eq:explicitA}) represent a nearest-neighbor hopping process in the auxiliary space ${\cal H}$, they can -- for any fixed chain length $n$ -- be truncated to a
finite $d=1+\lfloor n/2 \rfloor$ dimensional Hilbert space ${\cal H}_d$ spanned by $\{\ket{0},\ket{1}\ldots\ket{d-1}\}$, still making the expressions (\ref{eq:MPS},\ref{eq:matS},\ref{eq:LTN}) exact.
(v) Since hopping amplitudes (\ref{eq:explicitA},\ref{eq:a}) are all {\em linear functions} of the coupling $\varepsilon$, the un-normalized NESS density operator $R$ is a {\em polynomial} in $\varepsilon$ of degree not larger than $2n$. In fact the degree is $2n-2$ as easily checked by explicit computation.
(vi) LTN (\ref{eq:LTN}) can be understood as an MPO on a tensor product auxiliary space ${\cal H}\otimes {\cal H}$, namely 
\be
R = \!\!\sum_{\un{s}\in\{0,\pm,{\rm z}\}^n}\!\!\!
\bra{0}\!\otimes\!\bra{0} \mm{B}_{s_1}\mm{B}_{s_2}\cdots \mm{B}_{s_n}\ket{0}\!\otimes\!\ket{0}\prod_{j=1}^n \sigma^{s_j}_j,
\label{eq:MPR}
\ee
introducing effectively $d^2 \times d^2$ dimensional matrices
\be
\mm{B}_s = (\tr\,{\sigma^s}^\dagger\sigma^s)^{-1}\!\!\!\!\sum_{\nu,\nu',\mu\in\{0,1\}}\!\!\! \sigma^s_{\nu',\nu} \mm{A}_{\mu-\nu'}\otimes\bar{\mm{A}}_{\mu-\nu}.
\label{eq:Bs}
\ee

{\em Computation of observables.-} Eq. (\ref{eq:MPR}) is a starting point for computation of expectations of physical observables
$\ave{A} = \tr\rho_\infty A = \tr R A/\tr R$. The normalization constant is computed as 
$\tr R = 2^n\bra{0}\!\otimes\!\bra{0} {\mm{B}_0\!}^n \ket{0}\!\otimes\!\ket{0}$, and
a general expectation of a Pauli operator product reads
$
\ave{\prod_{j=1}^n {\sigma^{s_j}_j}^\dagger} =
\frac{\bra{0}\!\otimes\!\bra{0} \mm{B}_{s_1}\cdots\mm{B}_{s_n}\ket{0}\!\otimes\!\ket{0}}{\bra{0}\!\otimes\!\bra{0} {\mm{B}_0\!}^n \ket{0}\!\otimes\!\ket{0}}  \prod_{j=1}^n \frac{\tr{\sigma^{s_j}}^\dagger\sigma^{s_j}}{2}.
$
For observables which are only products of $\sigma^{\rm z}_j$, say magnetization profile $\ave{\sigma^{\rm z}_j}$, spin-spin correlations $\ave{\sigma^{\rm z}_j \sigma^{\rm z}_k}$, etc., one can use the same trick as in \cite{new} to further simplify the calculations.
Namely, $\mm{B}_0$ and $\mm{B}_{\rm z}$ leave the auxiliary subspace ${\cal K}$ of diagonal vectors, spanned by $\{ \ket{r}\otimes\ket{r},r=0,1,2,\ldots\}$, invariant, $\mm{B}_{0,\rm z} {\cal K} \subseteq {\cal K}$. As the initial (final) vector $\ket{0}\otimes\ket{0}$ is also a member of ${\cal K}$, we can 
reduce the domain of our operators to ${\cal K}$ defining tridiagonal transfer matrices (TMs), $\mm{T} := \mm{B}_0|_{\cal K}, \mm{V} := \mm{B}_{\rm z}|_{\cal K}$, or explicitly -- using identification $\ket{r}\otimes\ket{r}\to \ket{r}$
\begin{eqnarray}
\mm{T} &=& \sum_{r=0}^\infty\Bigl(
\left|a^0_r\right|^2 \ket{r}\bra{r}  +  \frac{\left|a^+_r\right|^2}{2} \ket{r}\bra{r\!+\!1} +  \frac{\left|a^-_r\right|^2}{2} \ket{r\!+\!1}\bra{r}\Bigr), \nonumber \\
\mm{V} &=& \sum_{r=0}^\infty\Bigl(
\frac{\left|a^+_r\right|^2}{2} \ket{r}\bra{r\!+\!1}-
\frac{\left|a^-_r\right|^2}{2} \ket{r\!+\!1}\bra{r} 
\Bigr), \label{eq:TMdef}
\end{eqnarray}
where we supplement (\ref{eq:a}) by $a^0_0:=1$, $a^+_0:=\ii \varepsilon$, $a^-_0:=1$,
so that the physical observables are computed in terms of $d \times d$ matrix products
\begin{eqnarray}
\ave{\sigma^{\rm z}_j} &=& \bra{0}\mm{T}^{j-1}\mm{V}\mm{T}^{n-j}\ket{0}/\bra{0}\mm{T}^n\ket{0}, \label{eq:expz} \\
\ave{\sigma^{\rm z}_j \sigma^{\rm z}_k} &=& \bra{0}\mm{T}^{j-1}\mm{V}\mm{T}^{k-j-1}\mm{V}\mm{T}^{n-k}\ket{0}/\bra{0}\mm{T}^n\ket{0}, \; {\rm etc.} \nonumber
\end{eqnarray}
Another class of interesting physical observables are the spin current $J_j = \ii (\sigma^+_j \sigma^-_{j+1} - \sigma^-_j \sigma^+_{j+1})$, local 
energy $h_j$, or similar, which can be all formulated in terms of expectations of a non-Hermitian one-sided hopping operator
$w_j := \sigma^-_j \sigma^+_{j+1}$. The product $\mm{B}_+ \mm{B}_-$ also leaves the diagonal space ${\cal K}$ invariant, so we introduce another
vertex operator $\mm{W} :=  \frac{1}{4} \mm{B}_+ \mm{B}_-|_{\cal K}$, or explicitly
\begin{eqnarray}
&&\mm{W} = \frac{1}{4}\sum_{r=0}^\infty \Bigl\{ 
a^0_r \bar{a}^0_{r+1} \bigl( \left|a^+_r\right|^2 \ket{r}\bra{r\!+\!1} +   \left|a^-_r\right|^2 \ket{r\!+\!1}\bra{r}\bigr)  \nonumber \\
&&+ (a^0_r)^2 \bar{a}^+_r \bar{a}^-_r \ket{r}\bra{r} + a^+_r a^-_r (\bar{a}^0_{r+1})^2 \ket{r\!+\!1}\bra{r\!+\!1}\Bigr\}, \quad \label{eq:defW} 
\end{eqnarray}
in terms of which the hopping expectation reads as
\be
\ave{w_j} = \bra{0}\mm{T}^{j-1} \mm{W} \mm{T}^{n-j-1}\ket{0}/\bra{0}\mm{T}^n\ket{0}.
\label{eq:w}
\ee
Eqs. (\ref{eq:TMdef},\ref{eq:defW},\ref{eq:a}) imply ${\rm Im}\mm{W} = -\frac{\varepsilon}{4}\mm{T}$ so the spin current $\ave{J_j} = -2{\rm Im}\ave{w_j}$ is independent of the position $j$, manifesting local conservation law of magnetization.

\begin{figure}
          \centering	
	\includegraphics[width=1.05\columnwidth]{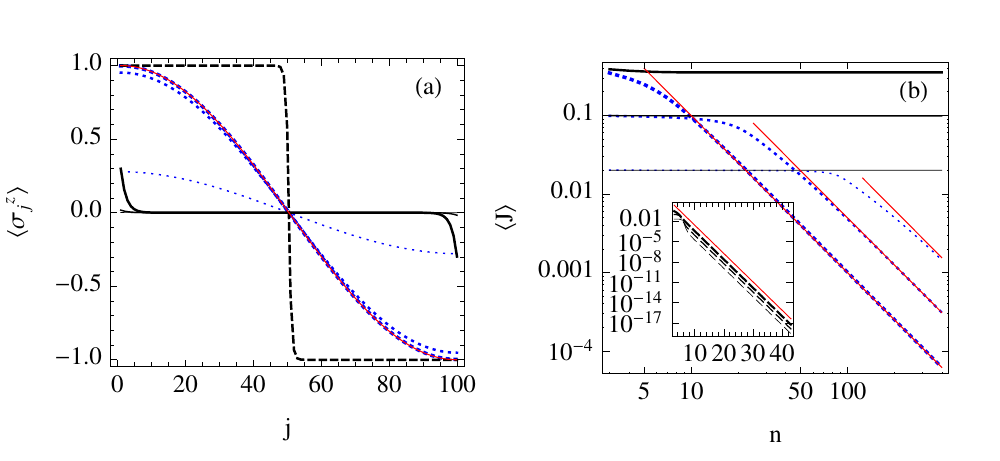}
	\vspace{-6mm}
	
	\caption{(color online).
	Spin profiles $\ave{\sigma^{\rm z}_j}$ at $n=100$ (a), and spin currents $\ave{J}$ vs. size $n$ (b), for $\Delta=3/2$ (dashed), $\Delta=1$ (dotted/blue), 
	$\Delta=1/2$ (full curves), all for three different couplings $\varepsilon=1,1/5, 1/25$ using thick, medium, thin curves, respectively. 
	Red full curves show closed-form asymptotic results  [see text]:
	$\ave{\sigma^{\rm z}_j}=\cos \pi \frac{j-1}{n-1}$, $\ave{J} =\pi^2 \varepsilon^{-1} n^{-2}$ for $\Delta=1$ in the main panels (a,b),
	and  $\ave{J} \propto e^{-n \,{\rm arcosh}\Delta}$ in (b)-inset.}
 	\label{fig:results}
\end{figure}

Let us now discuss some explicit results, graphically summarized in Fig.~\ref{fig:results}. We note that formulae (\ref{eq:expz},\ref{eq:w}) give efficient computational prescription which yields any observable of this type in ${\cal O}(n^2)$ arithmetic operations. In order to ensure numerical stability and to avoid singularities
we suggest to choose the signs $\tau_k$ in computation of auxiliary hopping amplitudes (\ref{eq:a}) as $\tau_k = 1$ for $\cos k \lambda \ge 0$, and $\tau_k = -1$ for $\cos k \lambda < 0$. 
For certain values of parameters even closed form results can be obtained. Analogously to the perturbative case \cite{new}, TMs have effective finite rank $m+1$, i.e. they close on ${\cal H}_{m+1}$, for a dense set of anisotropies,
$\lambda = \pi l/m$, which densely cover the easy-plane regime $|\Delta| < 1$. This happens because then $a^+_m = 0$ for odd $m$, or $a^-_m = 0$ for even $m$, and the auxiliary hopping process gets cut. For example, for $\Delta = 1/2=\cos \pi/3$, we calculate spin profiles and currents by
iterating a reduced TM $\mm{T}' = \mm{T}|_{{\cal H}_3}$
\be
\mm{T}' = \begin{pmatrix}
1 & \varepsilon^2/2 &  0 \cr
1/2 & (1+\varepsilon^2)/4 & (9+\varepsilon^2)/24 \cr
0 & 3(1+\varepsilon^2)/8 & (1+\varepsilon^2)/4 
\end{pmatrix}
\ee
combined with the reduced $3\times 3$ vertex matrices $\mm{V}'=\mm{V}|_{{\cal H}_3}$,$\mm{W}'=\mm{W}|_{{\cal H}_3}$, 
Explicit expressions for $\ave{\sigma^{\rm z}_j}, \ave{\sigma^{\rm z}_j \sigma^{\rm z}_k}, \ave{J_j}$ can easily be obtained by means of diagonalization of $\mm{T}'$. We obtain exponential convergence towards the thermodynamic limit (TL), $n\to\infty$, with the rate given by the
ratio of two leading eigenvalues of $\mm{T}'$, and asymptotically flat spin profiles $\ave{\sigma^{\rm z}_j}\approx 0$ (Fig.~\ref{fig:results}a). We prove ballistic transport by explicitly computing the limit
$\ave{J_j}|_{n\to\infty} =  \frac{\left(\sqrt{ 81 + 74 \varepsilon^2 + 9 \varepsilon^4} - 7 - 3 \varepsilon^2 \right)\varepsilon}{4(1+\varepsilon^2)}$ (Fig.~\ref{fig:results}b), 
having a non-monotonic $\varepsilon$-dependence starting as $\sim \varepsilon/2$ for small $\varepsilon$ (consistent with \cite{new}), a maximum at $\varepsilon^*\approx 1.63$, and decaying asymptotically as $\sim 4/(3\varepsilon)$ for large $\varepsilon$, qualitatively agreeing with similar
results for the non-interacting XX \cite{karevski} and XY chains \cite{bojan}.
Similar finite dimension analysis can be made for some larger denominators $m$. On the other hand, for $\Delta \ge 1$, the TM $\mm{T}$ has always an infinite rank. In the easy-axis regime $|\Delta| > 1$, explicit computations reveal almost $\varepsilon$-independent kink-shaped spin density profile (Fig.~\ref{fig:results}a) -- agreeing with numerical simulations of negative differential conductance  \cite{giuliano} -- and asymptotically exponentially decaying current, $\ave{J_j} \propto (|\Delta| + \sqrt{\Delta^2-1})^{-n}$ (Fig.~\ref{fig:results}b-inset), consistent with suggested ideally insulating behavior \cite{marcin}.

At the end, let us briefly focus on the isotropic case $\Delta=1$.
In this case, our hopping matrices (\ref{eq:explicitA}) have to be regularized by taking $\tau_{2k-1}=1$, $\tau_{2k}=-1$, and $c = 1/\lambda$ before taking the limit $\lambda\to 0$, yielding the hopping amplitudes:
$a^0_r=1+\ii \varepsilon r/2$, $a^+_0=\ii\varepsilon$, $a^+_{2k-1}=2k + \ii \varepsilon k (k-\frac{1}{2})$, $a^+_{2k}=2k + \ii \varepsilon k^2$, $a^-_0=1$, $a^-_{2k-1}=\ii\varepsilon$, $a^-_{2k}=\ii \varepsilon (k+\frac{1}{2})/k$.
The following formulae can be verified with some effort
\begin{eqnarray}
\!\!&&[\mm{T},[\mm{T},\mm{V}]] =-\frac{\varepsilon^2}{4}(2\mm{V}+\{\mm{T},\mm{V}\}), \label{eq:secder}\\
\!\!&&\bra{0}(\mm{T}-\mm{V})=\bra{0},\quad (\mm{T}+\mm{V})\ket{0}=\ket{0},
\label{eq:bcTV}\\
\!\!&&\frac{\bra{0}\mm{T}^{n}\ket{0}}{\bra{0}\mm{T}^{n-1}\ket{0}} \simeq \varepsilon^2 \biggl({ \frac{(4n-3)^2}{32\pi^2}} - \alpha\biggr)+1+ {\cal O}(n^{-1}),\;\;\;\;\label{eq:Tlead} 
\end{eqnarray}
where $\alpha \approx 0.0346$.
Multiplying (\ref{eq:secder}) by $\bra{0}\mm{T}^{j-1}$ from the left, and $\mm{T}^{n-j-2}\ket{0}$ from the right, and using (\ref{eq:Tlead}) we obtain in the continuum limit
 $M(x\equiv \frac{j-1}{n-1}):=\ave{\sigma^{\rm z}_j}$ a differential equation $M''(x)=- \pi^2 M(x) + {\cal O}(\frac{1}{n})$, and from (\ref{eq:bcTV}) the boundary conditions $M(0)=-M(1)=1+{\cal O}(\frac{1}{\varepsilon^2 n^2})+{\cal O}(\frac{1}{n})$,
yielding a magnetization profile $M(x) = \cos \pi x$, or $\ave{\sigma^{\rm z}_j} \simeq \cos \pi \frac{j-1}{n-1}$,
for {\em arbitrary} $\varepsilon \gg \varepsilon^* =  2\pi/n$ (Fig.~\ref{fig:results}a).
Similarly we use (\ref{eq:secder}-\ref{eq:Tlead}) and the continuum approximation
to calculate the connected correlator $C(x\equiv\frac{j-1}{n-1},y\equiv\frac{k-1}{n-1}):=\ave{\sigma^{\rm z}_j \sigma^{\rm z}_k} - \ave{\sigma^{\rm z}_j}\ave{\sigma^{\rm z}_k}$, for $j\neq k$.
However, as it turns out that the leading order ${\cal O}(n^0)$ of $C(x,y)$ exactly vanishes, we solve the corresponding differential equations perturbatively in the next order in $1/n$.
Straightforward but tedious calculation gives 
$C(x,y) \simeq \frac{\pi}{4n}f({\rm min}(x,y),{\rm max}(x,y))+{\cal O}(\frac{1}{n^2})$, where $f(x,y)= 2 \pi  x (y-1) \sin (\pi  x) \sin (\pi  y)+\cos (\pi  x) ((1-2 y) \sin (\pi  y)+\pi  (y-1) y \cos (\pi  y))$.
This is another, now analytic, indication of {\em long-range correlations} in far from equilibrium quantum NESS recently observed numerically or in non-interacting systems \cite{ProZni2010}.
Eq. (\ref{eq:Tlead}) and ${\rm Im\,}\mm{W} = -\frac{\varepsilon}{4}\mm{T}$ imply anomalous sub-diffusive scaling $\ave{J_j}=\frac{\varepsilon}{2} \bra{0}\mm{T}^{n-1}\ket{0}/\bra{0}\mm{T}^n\ket{0} 
\approx \pi^2 \varepsilon^{-1} n^{-2}$, again valid for any $\varepsilon \gg \varepsilon^*(n)$ (Fig.~\ref{fig:results}b).
For $\varepsilon \ll \varepsilon^*$ we reproduce the perturbative result \cite{new}, $\ave{J_j} =\frac{1}{2}\varepsilon$, $\ave{\sigma^{\rm z}_j}=\frac{1}{4}\varepsilon^2 (n+1-2j)$.

{\em Discussion.-} An explicit LTN/MPO ansatz has been written describing the many-body density matrix of NESS of strongly boundary driven XXZ chain, for any bath-coupling strength. Computation of the physical observables in NESS is facilitated in terms of tridiagonal transfer matrices which are reminiscent -- except for non-conservation of `probability' -- of a classical Markov process in the auxiliary space.
Results in TL can be obtained by studying the spectral properties of the transfer operator.
Studying Liouvillian gap or relaxation rates to NESS and related uniqueness of NESS is yet to be addressed.
It also remains open to what extend our solution (\ref{eq:defR}-\ref{eq:a}) can be generalized to other bath-models, for example the case of weak driving has fundamentally different physical properties  \cite{pz,markonew}. Our method seems to open a new ground for constructing exactly solvable nonequilibrium quantum problems in one dimension, and seems to be unrelated \cite{foot} to existing algebraic methods \cite{blythe,hosho}. 
New exactly solvable models could perhaps be constructed by studying alternative cubic algebras of type (\ref{eq:algebra}). 

Discussions with M. \v Znidari\v c and support by the grants J1-2208 and P1-0044 of ARRS (Slovenia) are acknowledged.


\begin{thebibliography}{10}
\bibitem{temperature} Assuming that the magnetic baths are in grand-canonical states, their temperatures are infinite.
\bibitem{foot} We have shown explicitly that no `integration-constant' operators exist such that they would together with $\mm{B}_s$ (\ref{eq:Bs}) satisfy a quadratic algebra, like in all known solvable classical exclusion processes \cite{blythe}.
\end{thebibliography}
\end{document}